\documentclass{article}
\usepackage{ijcai25}

\usepackage{times}
\usepackage{soul}
\usepackage{url}
\usepackage[hidelinks]{hyperref}
\usepackage[utf8]{inputenc}
\usepackage[small]{caption}
\usepackage{graphicx}
\usepackage{amsmath}
\usepackage{amsthm}
\usepackage{booktabs}
\usepackage{algorithm}
\usepackage{algorithmic}
\usepackage[switch]{lineno}
\usepackage{amssymb,amsfonts,mathrsfs,dsfont,mathtools}
\usepackage{comment}

\title{Deep Reinforcement Learning Algorithms for Option Hedging}

\author{
Andrei Neagu$^1$
\and
Frédéric Godin$^2$\and
Leila Kosseim$^{1}$
\affiliations
$^1$Dept. of Computer Science and Software Engineering, Concordia University, Montréal, Canada\\
$^2$Dept. of Mathematics \& Statistics, Concordia University, Montréal, Canada\\
\emails
\{andrei.neagu, frederic.godin, leila.kosseim\}@concordia.ca
}
\begin{document}

\maketitle

\begin{abstract}

Dynamic hedging is a financial strategy that consists in periodically transacting one or multiple financial assets to offset the risk associated with a correlated liability. Deep Reinforcement Learning (DRL) algorithms have been used to find optimal solutions to dynamic hedging problems by framing them as sequential decision-making problems. However, most previous work assesses the performance of only one or two DRL algorithms, making an objective comparison across algorithms difficult. In this paper, we compare the performance of eight DRL algorithms in the context of dynamic hedging; Monte Carlo Policy Gradient (MCPG), Proximal Policy Optimization (PPO), along with four variants of Deep $Q$-Learning (DQL) and two variants of Deep Deterministic Policy Gradient (DDPG). Two of these variants represent a novel application to the task of dynamic hedging. In our experiments, we use the Black-Scholes delta hedge as a baseline and simulate the dataset using a GJR-GARCH(1,1) model. Results show that MCPG, followed by PPO, obtain the best performance in terms of the root semi-quadratic penalty. Moreover, MCPG is the only algorithm to outperform the Black-Scholes delta hedge baseline with the allotted computational budget, possibly due to the sparsity of rewards in our environment. 

All datasets and code are publicly available \href{https://github.com/Andrei-T-Neagu/DRL_in_Finance}{here}.
\end{abstract}

\section{Introduction}

Hedging is a financial practice in which one or multiple assets are traded to minimize the risk associated with a correlated financial liability. Dynamic hedging is an approach where the hedging portfolio is periodically rebalanced, as opposed to static hedging where the hedging position is set at the initiation and left unchanged until the expiry date of the liability. Dynamic hedging has the potential to be more effective at minimizing risk as it allows constant readjustments of the hedging portfolio's composition to account for the changing risk profile of the liability being hedged. The problem of dynamic hedging can be framed as a sequential decision-making problem; thus, DRL techniques can be applied to find an optimal solution. Under this framework, the states at each time step represent the market conditions and the actions correspond to the number of hedging asset shares to be included in the hedging portfolio.

Most of the previous related works only investigate the performance of one or two DRL algorithms for the task of dynamic hedging, making an objective comparison between algorithms difficult.

\subsection{Contributions}
\label{se:contributions}

The main contributions of this paper are two-fold: \textbf{1)} We provide a standard comparison of the performance and time efficiency of the most widely used DRL algorithms in the literature to tackle dynamic hedging optimization. \textbf{2)} We evaluate and analyze the performance of variants of these DRL algorithms not previously explored in the DRL dynamic hedging literature, namely: Dueling DQL and Dueling Double DQL. 

\section{Related Work}

The first papers tackling dynamic hedging using DRL are \cite{BuehlerDeepHedging} and \cite{Halperin}. The former seminal paper, along with some subsequent works, uses a Monte Carlo Policy Gradient (MCPG) algorithm, see for instance
\cite{BuehlerDeepHedging,CarbonneauGodinERP,horvath2021deep,CarbonneauGodin2021DeepEqualRiskPricingMultiple,NeaguDeepHedging}. This approach directly learns the optimal policy, which is the mapping from states to actions that minimizes a given risk measure applied to the terminal hedging loss.
Since then, multiple papers have expanded upon this work by using different DRL algorithms. 
\cite{DuReplication} have used \textit{value-based} algorithms, such as Deep $Q$-Learning (DQL) \cite{MnihDQN}, which learn the value of taking any action in a given state; the optimal policy is then derived by taking the action with the highest value in each state. 
Moreover, the Fitted Q-Iteration algorithm, which is an adaptation of Q-learning, is used in the QLBS model of \cite{Halperin}. The performance of the QLBS approach is investigated in \cite{stoiljkovic2025advanced}.
Other papers used \textit{actor-critic} algorithms which combine policy- and value-based approaches, see for instance \cite{CaoGammaAndVega,MarzbanExpectile,SharmaHedging}. Actor-critic algorithms considered include Proximal Policy Optimization (PPO) \cite{Schulman2017PPO} used by \cite{DuReplication}, Deep Deterministic Policy Gradient (DDPG) \cite{Lillicrap2016continuous} used by \cite{CaoDeepHedging} and Twin Delayed DDPG (TD3) \cite{Fujimoto2018TD3} used by \cite{Mikkila2022TD3}. However, all these works focus on the evaluation of a single algorithm, making standard comparisons difficult. To our knowledge,  \cite{DuReplication} is the only work that analyzes and compares the performance of multiple algorithms, but limits this analysis to only two, namely: DQL and PPO.

\section{Background}

\begin{table*}[ht]
\centering
\begin{tabular}{| r  l || r  l || r  l |}
\toprule
\toprule
\multicolumn{6}{|c|}{Financial Symbols}\\
\midrule
$\mathbf{t}$ & \textbf{time step} & $X$ & hedging strategy (\# stock shares) & $K$ & strike price\\
$T$ & expiry & $p_0$ & option premium & $\rho$ & risk measure\\
$R$ & random loss variable & $\mathcal{P}_X$ & profits & $M$ & cash reserve\\
$r_f$ & risk-free rate & $e^{r_f}$ & accrual factor & $c$ & transaction total\\
$\mathbf{S}$ & \textbf{underlying stock price} & $\mathbf{V}$ & \textbf{hedging portfolio value} & $\delta_t$ & time step length\\
\midrule
\midrule
\multicolumn{6}{|c|}{DRL Symbols}\\
\midrule
$s$ & state & $a$ & action & $r$ & reward\\
$\pi(a|s)$ & stochastic policy & $\pi(s)$ & deterministic policy & $\gamma$ & discount factor\\
$\alpha$ & policy learning rate & $\beta$ & value function learning rate & $\theta$ & policy parameters\\
$\phi$ & value function parameters & $Q(s,a)$ & state-action value function & & \\
% $A(s,a)$ & advantage function & $\hat{G}$ & empirical return & $\mathcal{D}$ & experience memory buffer\\
\bottomrule
\end{tabular}
\caption{Symbols and their definition, with state variables in bold.}
\label{tb:notations}
\end{table*}

This section provides financial background knowledge, describes the market environment model considered, and provides an overview of the different DRL algorithms that are compared. For the reader's convenience, all financial and DRL notations used in this paper are summarized in Table~\ref{tb:notations}.

\subsection{Financial Background} 
\label{se:FinancialBackground}
An option is a financial contract which gives the right to purchase (for a \textit{call} option) or to sell (for a \textit{put} option) an asset, referred to as the underlying asset, at a given date $T$, called the \textit{expiry} and at a predetermined price, called the \textit{strike price} $K$. Movements in the price of the underlying asset cause either an increase or a decrease in the value of the option. This correlation can be leveraged by a financial institution which issued the option; the value of which is a liability to them. To offset the risk of a potential appreciation or depreciation in the value of the option, the financial institution can periodically purchase or sell shares of the underlying asset.

In our work, we hedge the sale of a call option, for which we receive a cash amount, called a \textit{premium} $p_0$, and hedge the risk of the option value increasing by periodically adjusting the number of shares of the underlying stock held at each time step $t=0,1,\dots,T$.

\subsubsection{Dynamic Hedging as an Optimization Problem} To offset potential losses related to the sale of the call option at the expiry $T$, a hedging portfolio consisting of cash and underlying stock shares is set up.
The dynamic hedging problem consists of selecting a hedging strategy $X=\{X_t\}^T_{t=1}$ which minimizes the possible risk at the expiry $T$, where $X_t$ corresponds to the number of stock shares in the portfolio during $(t-1,t]$. That is, we wish to solve
\begin{equation}
\label{globhedging}
X^*=\underset{X}{\arg\min} \, \rho \left(R\right)
\end{equation}
where $\rho$ is a risk measure mapping a random loss variable $R$ into a real number representing the risk, and where ${R=-\mathcal{P}_X}$ is the negative of the total profits $\mathcal{P}_X$ at time $T$ under the hedging strategy $X$. Several risk measures $\rho$ have been used in the literature \cite{CarbonneauGodin2021DeepEqualRiskPricingMultiple}. In our work, we consider the \textit{root semi-quadratic penalty (RSQP)} risk measure 
\begin{equation}
\label{eq:RSQP}
    \rho^{RSQP}(R)=\sqrt{\mathbb{E}\left[ R^2\mathds{1}_{ \{R>0\} }    \right]}.
\end{equation}
where $\mathds{1}_{\{R>0\}}$ is an indicator variable taking value $1$ if $R>0$ and $0$ otherwise.

The popular \textit{quadratic penalty} is not considered as a risk measure, since it has the downside of penalizing gains, unlike its \textit{semi-quadratic} counterpart.

Let the quantity $M_t$ denotes the amount of cash in the portfolio at time $t$ right before any transaction. Since all transactions of stock shares are financed through the portfolio cash reserve represent by process $M$, the hedging portfolio is said to be \textit{self-financing}.
At each time period, the cash reserve is set to accrue (increase) by a factor of $e^{r_f}$ where $r_f$ is the one-period risk-free rate.
The cash amount in the portfolio can be found recursively through
\begin{eqnarray}
M_t = 
\begin{cases}
p_0 & \text{ for } t=0,
\\ (M_{t-1} - c_{t}(X))e^{r_f} & \text{ for } t=1,\ldots,T,
\end{cases}
\end{eqnarray}
with the stock transaction total amount at time $t$ being
\begin{equation}
c_t(X)=
\begin{cases}
0 & \text{ for } t=0,\\
S_{t-1} (X_{t} - X_{t-1}) & \text{ for } t=1,\ldots,T,
\end{cases}
\end{equation}
where $S_t$ is the underlying stock price at time $t$ and $X_0 = 0$.
Consider an agent hedging the sale of a call option. If, at the expiry $T$, the underlying stock price is larger than the strike price $K$, the buyer of the option will choose to exercise their right to buy the underlying stock at the strike price $K$. After implementing the hedging strategy $X$, the total profit for the agent right after the expiry is:
\begin{equation}
\mathcal{P}_X = S_T X_T + M_T - \mathds{1}_{E} (S_T-K),
\label{eq:profits}
\end{equation}
with $\mathds{1}_{E}$ being the indicator variable worth $1$ if event $E \equiv\{S_T>K\}$ occurs, or $0$ otherwise.

An important value that we use as a state variable into the DRL algorithm is the portfolio value:

\begin{equation}
    V_t = S_t X_t + M_t,
\end{equation}
which is simply the sum of the value of the underlying stock shares currently held and the cash amount at time $t$.

\subsection{Market Environment}
We define an underlying stock price process $S = \{S_t\}^T_{t=0}$. Set $S_t = S_{t-1}\exp(Y_{t})$, where $Y_t$ is the time-$t$ log-return of the underlying stock. Log-returns are modeled with the GJR-GARCH(1,1) model \cite{Glosten1993GARCH}:
\begin{align}
\label{eq:gjr-garch}
    Y_t &= \mu + \varepsilon_t \\
    \varepsilon_t &= \sigma_t z_t \nonumber\\
    \sigma^2_t &= \nu_0 + (\nu + \lambda I_{t-1})\varepsilon^2_{t-1} + \xi\sigma^2_{t-1} \nonumber\\
    \intertext{with}
    I_{t-1}=&\begin{cases}
        0 & \text{ if } Y_{t-1}\geq\mu\\
        1 & \text{ if } Y_{t-1}<\mu\\
    \end{cases}\nonumber
\end{align}
where $\{\sigma^2_t\}^T_{t=1}$ are the conditional variances of log-returns, $\{\mu,\lambda\}\in\mathbb{R}$ and $\{\nu_0,\nu,\xi\}\in\mathbb{R}^+$ are the model parameters, and $\{\varepsilon_t\}^T_{t=1}$ is a series of standard normal random variables.

The GJR-GARCH(1,1) model expands upon the geometric Brownian motion from the classic Black-Scholes model \cite{BSM1973} by assuming the presence of stochastic volatility, volatility clustering and the leverage effect; features that more closely resemble real stock prices behaviour.

\subsection{Deep Reinforcement Learning Algorithms}
% PG hyperparams: state_size, num_layers, hidden_size, gamma, lr, batch_size, num_updates, lr_scheduler, early stopping
% DQL hyparams: state_size, min_action, max_action, action_delta, num_layers, hidden_size, gamma, lr, batch_size, num_updates, lr_scheduler, early stopping, epsilon, espilon_min, memory_size, loss, epsilon_decay
% Double DQL hyparams: same, tau, target_update
% Dueling DQL hyparams: same
% Dueling Double DQL hyparams: same, tau, target_update
% PPO hyperparams: state_size, num_layers, hidden_size, state_size2, num_layers2, hidden_size2, gamma, lr, lr2, batch_size, num_updates, lr_scheduler, lr_scheduler2, early stopping, epochs, clip_eps, loss
% DDPG hyperparams: state_size, num_layers, hidden_size, state_size2, num_layers2, hidden_size2, gamma, lr, lr2, batch_size, num_updates, lr_scheduler, lr_scheduler2, early stopping, memory_size, loss, tau, target_update, epsilon, min_action, max_action, epsilon_decay
% TD3 hyperparams: same, state_size3, num_layers3, hidden_size3, lr3, lr_scheduler3
\begin{table}[ht]
    \centering
    \begin{tabular}{l|l|p{1.6cm}|r}
        \toprule
        \toprule
        Algorithm & Type & Action Space & \multicolumn{1}{p{1cm}}{\# of Hyperparameters}\\
        \midrule
        \midrule
        DQL & Value-based & Discrete & 17\\
        \quad Double DQL & Value-based & Discrete & 19\\
        \quad Dueling DQL* & Value-based & Discrete & 17\\
        \quad DD DQL* & Value-based & Discrete & 19\\
        MCPG & Policy-based & Continuous & 9\\
        PPO & Actor-critic & Continuous & 17\\
        DDPG & Actor-critic & Continuous & 22\\
        \quad TD3 & Actor-critic & Continuous & 27\\
        \bottomrule
    \end{tabular}
    \caption{Characteristics of the DRL algorithms explored in this paper. The algorithms indicated by stars are the two variants not previously explored in the dynamic hedging literature.}
    \label{tab:DRL_algorithms}
\end{table}

Our work compares the performance of eight DRL algorithms. Table~\ref{tab:DRL_algorithms} describes the attributes of each algorithm as well as showing which variants stem from which baseline algorithm (i.e. TD3 from DDPG; Double, Dueling, and Dueling Double (DD) DQL from DQL).

\subsubsection{Deep Q-Learning (DQL)}

Deep $Q$-Learning (DQL) (see \cite{MnihAtari}) has become a widely used DRL algorithm since it was first used to surpass the performance of a human player on the suite of Atari games in 2013. DQL works by deriving the optimal policy from an action-value function which is derived from the \textit{Bellman equation}. The optimal action-value function is the expected return (sum of rewards $r$) when taking a given action $a$ in state $s$ and following the optimal policy $\pi^*$ subsequently: \mbox{$Q^*(s,a)=\mathbb{E}\left[ \sum^\infty_{u=t} \gamma^{u-t} r_u | s_t\!=\!s,a_t\!=\!a,(a_{u+1},)^\infty_{u=t} \!\sim\! \pi^*\right]$} with $\gamma$ being a discount factor. 

% DQL solves the recursion
% \begin{equation*}
%     Q^*(s,a)=\mathbb{E}_{s_{t+1}}\!\Big[r_t\!+\!\gamma\max_{a'}Q^*(s_{t+1},a')|s_t\!=\!s,a_t\!=\!a\Big].
% \end{equation*}
% which is used by Value-based RL approaches. The optimal actions in a state $s$ is that maximizing the $Q$-function: $a^*(s) = \argmax Q^*(s,a)$.
% %where $s$ and $s'$ are the current and next states, $a$ and $a'$ are the current and next actions,  %and $\mathcal{S}$ and $\mathcal{A}$ are the state and action spaces.

In DQL, the optimal action-value function is represented with a neural network with parameter set $\phi=\{\phi_i\}^{\mathcal{I}-1}_{i=0}$ which is estimated iteratively, for a total of $\mathcal{I}$ iterations. 
% Denote by $\phi_i$ the parameter set at iteration $i$. The training involves another version of the parameter set $\bar{\phi}$ called target parameters. $\bar{\phi}$ fluctuates more slowly to enhance stability and is updated using Polyak averaging: $\Bar{\phi}_{i+1}\gets\Bar{\beta}\phi_{i}+(1-\Bar{\beta})\Bar{\phi}_{i}$, where $\Bar{\beta}$ is the target parameters' learning rate. At iteration $i$, a mini-batch of $N$ transitions $(s,a,r,s')\sim\mathcal{D}$ is sampled randomly from an experience memory buffer $\mathcal{D}$. Updates to the action-value function are then performed using stochastic gradient descent
% \begin{equation}
%     \phi_{i+1}\gets\phi_{i}-\beta\nabla_{\phi_i}L_i(\phi_i)
% \end{equation}
% where $\beta$ is the learning rate and
% \begin{align*}
%     L_i(\phi_i)=\!
%     \frac{1}{N}\!\!\sum^N_{n=1}\!\!\Big(r_{n} \!\!+\!\! \gamma \max_{a'_n} Q\big(s'_{n},a'_n;\Bar{\phi}_{i}\big)\!-\!Q\big(s_{n},a_{n};\phi_i\big)\Big)^2    
% \end{align*} 
% with $s'_n$ $a'_n$ being sampled successor state and action in transition $n$, respectively.

%When using a neural network as the function approximator, $Q$-Learning is called Deep $Q$-Learning (DQL).

A common problem with DQL is that it tends to overestimate the value of state-action pairs. To counter this, different variations have been proposed such as Double DQL \cite{vanHasselt2015DoubleDQL}, which uses two different value functions, one to pick the action maximizing the state-action value function estimate and one to evaluate the value of that action. Another variant of DQL is Dueling DQL \cite{Wang2016DuelingDQL} which splits the state-action value function $Q$ into two parts: a state value function, and an advantage function which represents the relative value of taking a certain action compared to other actions in the same state. This leads to better generalization when actions have similar rewards and to better identification of states in which actions do not significantly affect the environment.

\subsubsection{Monte Carlo Policy Gradient (MCPG)}

Classic value-based RL methods tend to suffer from high computational complexity as we increase the dimension of the problem \cite{SuttonPGRL}. %Additionally, the estimate of the value function is biased.
An alternative approach to mitigate such problem is to directly parameterize the policy instead of deriving it from a value function \cite{SuttonPGRL}.%, which removes the bias but increases the variance of the iterative updates to the policy.

A simple instance of such approach is the Monte Carlo Policy Gradient found in \cite{BuehlerDeepHedging}. To improve the policy, multiple instances of the hedging loss $R$ are simulated based on the current policy. Its distribution is then used to assess the value of a given policy.
In our work, we employ a deterministic policy $a\sim\pi(s;\theta)$ as in \cite{BuehlerDeepHedging}. % and similarly to \cite{SilverDeterministicPolicyGradient}.
The policy parameters are updated iteratively using stochastic gradient descent with batches of $R\equiv\{R_n\}^N_{n=1}$ of $N$ i.i.d. loss variables %resulting from the transitions $\big(s_{(t,n)}, a_{(t,n)}, r_{(t,n)},s_{(t+1,n)}\big)$ where $a_{(t,n)}\sim\pi(s_{(t,n)};\theta_i)$ for all $t=0,\dots,T-1$ is 
simulated with the current policy parameters~$\theta_i$:
\begin{align}
\begin{split}
    \theta_{i+1}\gets\theta_{i}-\alpha\nabla_{\theta_i}\widehat{\rho}^{RSQP}(R)
\end{split}
\end{align}
where the gradient of the empirical estimate of the \textit{RSQP} risk measure $\widehat{\rho}^{RSQP}(R)$ is computed in closed-form with automatic differentiation packages. $\alpha$ is the learning rate of the procedure.

% \begin{algorithm}[H]
%     \caption{Policy Gradient (PG) with deterministic policy}\label{alg:reinforce}
%     \textbf{Parameters: } Learning rate $\alpha$, discount factor $\gamma$, number of time steps $T$, batch size $N$, initial random parameters $\theta_0$ of policy $\pi(s;\theta)$.\\
%     \textbf{Output: } An optimal deterministic policy $\pi^*(s;\theta)$.
    
%     \begin{algorithmic}[1]
%     \FOR{episode $=0,1,\dots,M$}
%     \STATE Initialize a batch of $N$ initial states $s_{(0,n)}$ by resetting the environment.
%     \FOR{$t=0, \dots, T-1$}
%         \STATE Take action $a_{(t,n)}\sim\pi(s_{(t,n)};\theta_i)$ and observe $r_{(t,n)}$, and $s_{(t+1,n)}$.
%         \STATE Compute $\hat{G}_{(t,n)}$ as an estimate to $Q(s_{(t,n)},a_{(t,n)})$.
%         \STATE Update the policy parameters using stochastic gradient ascent
%         % TO BE MODIFIED
%         \begin{align}
%         \begin{split}
%             \theta_{i+1}\gets\theta_{i}+\alpha\frac{1}{N}\sum^N_{n=1}\nabla_{\theta_i}\pi(s_{(t,n)};\theta_i)\\
%             \nabla_{a_{(t,n)}}Q(s_{(t,n)},a_{(t,n)}).
%         \end{split}
%         \end{align}
%     \ENDFOR
%     \ENDFOR
%     \end{algorithmic}
% \end{algorithm}

\subsubsection{Deep Deterministic Policy Gradient (DDPG)}

Deep Deterministic Policy Gradient (DDPG) (see \cite{Lillicrap2016continuous}) is an actor-critic algorithm. Instead of estimating the optimal policy as a by-product of the action-value function $Q$, it proposes a policy (the actor) that is progressively refined by using an estimated value function (the critic). 
% The action is selected as
% s\begin{equation}
%     a_t\sim\pi(s_t;\theta_t)+\nu
% \end{equation}
% where $\pi$ is a neural network with parameters $\theta$ acting as a deterministic policy and $\nu$ is a random variable sampled from a distribution $\mathcal{N}(0,\sigma)$ every time an action is taken and whose variance $\sigma$ converges to zero over time.

% While the policy $\pi$ is deterministic, the noise term $\nu$ allows the algorithm to conduct exploration. 
DDPG employs a deterministic policy $\pi(s;\theta)$, which reduces the sample complexity. Furthermore, the action-value function $Q(s,a)$ of the current policy is approximated using a neural network $Q(s,a;\phi)$ with parameters $\phi$. The objective of the DDPG algorithm is thus to estimate both actor and critic parameters $\theta$ and $\phi$, respectively. 
%The sampled action is therefore deterministic only asymptotically.

%Most RL algorithms in the literature make use of a stochastic policy.  However, this leads to pure exploitation (the action with the current highest value being taken) which leads to suboptimal actions being taken since the current action with the highest value might not be deemed optimal during the current iteration. Therefore, exploration (taking a random action in the hopes that it has a higher value) must be achieved through other means.

The DDPG algorithm has the advantage over the DQL algorithm of working naturally with continuous action spaces. Indeed, in DQL, we need to compute the action which has the maximum value $\max_a Q^*(s,a)$, which is feasible for finite discrete action spaces but difficult (or sometimes infeasible) for continuous action spaces. Conversely, $\pi$ is a differentiable function whose outputs lie in a continuous space.

However, DDPG also suffers from the same overestimation problem encountered in DQL \cite{Fujimoto2018TD3}. To mitigate this problem, Twin Delayed DDPG (TD3) \cite{Fujimoto2018TD3} was proposed. In TD3, two state action value functions $Q$ are learned and the smallest value of the two is used as the target to update the value function. Additionally, the target parameter updates are delayed, which is shown to improve the performance of the algorithm.

\subsubsection{Proximal Policy Optimization (PPO)}
\label{se:PPO}

The Proximal Policy Optimization (PPO) algorithm (see \cite{Schulman2017PPO}) has been developed with the objective of reducing the variance of the policy updates. It is an actor-critic algorithm which clips what is called the \textit{probability ratio}, a measure of how far the new policy is from the previous policy, and thus penalizes large updates to the policy so as to reduce the variance of updates and prevent the policy from performing catastrophic updates.

\section{Experimental Setting}
As indicated in Section~\ref{se:contributions}, our paper provides an objective comparison of multiple DRL algorithms in the context of dynamic hedging.
Specifically, we compare the performance and time efficiency of eight DRL algorithms: DQL, Double DQL, Dueling DQL, Dueling Double DQL, MCPG, DDPG, TD3, PPO; two of which have never been used in the previous literature to the best of our knowledge (Dueling DQL and Dueling Double DQL).

\subsection{Baseline}
\label{se:bsdh}
To evaluate our DRL model's optimal hedging strategies $X^*$, we compared them to the Black-Scholes delta hedge (B-S DH) \cite{BSM1973}, a commonly used hedging strategy baseline in the literature:
\begin{align}
    X_{t+1} &= \Phi(d_1), \quad\quad\quad \text{ for } t=0,\ldots,T-1, \label{eq:BSDH}
    \\d_1 &= \frac{\log(S_t/K) +(r_f+\sigma^2/2)(T-t)\delta_t}{\sqrt{\sigma^2(T-t)\delta_t}} \notag
\end{align}
where $\Phi$ is the standard normal cumulative distribution function, and $\delta_t$ is the time elapsing (in years) between any two time points $t$. In the absence of transaction costs, in continuous time, and for market dynamics following a geometric Brownian motion, this procedure is shown to completely eliminate risk \cite{BSM1973}, which explains its popularity. 

\subsection{Financial Setting}
In our experiments, we hedge the sale of a standard call option with strike price $K=100$ and one-year expiry with monthly time steps $\big(T=12$, $\delta_t=\frac{1}{12}\big)$ and employ the RSQP (Eq.~\ref{eq:RSQP}) as a risk measure. The risk-free rate is set to $r_f=0$.

\subsection{DRL Setting}
\label{se:DRL_Setting}

\subsubsection{Dataset}
To evaluate the performance of the different DRL algorithms, we trained them on simulated price paths following Eq.~\ref{eq:gjr-garch}, where the parameters of the GJR-GARCH(1,1) model \cite{Glosten1993GARCH} are learned through maximum likelihood estimates from monthly S\&P 500 prices from $2000/11/15$ to $2024/10/15$. These parameters are shown in Table~\ref{tab:garch_params}.
\begin{table}[ht]
    \centering
    \begin{tabular}{rl||rl||rl}
        \toprule
        \toprule
        $\mu$ & 0.00533410 & $\nu_0$ & 0.00018216 & $\nu$ & 0.00026564\\
        \midrule
        $\xi$ & 0.70611408 & $\lambda$ & 0.34275732 \\
        \bottomrule
    \end{tabular}
    \caption{Parameters of the GJR-GARCH model estimated from monthly S\&P 500 prices from $2000/11/15$ to $2024/10/15$.}
    \label{tab:garch_params}
\end{table}

We perform hyperparameter tuning for 200,000 updates on a validation set consisting of $2^{17}$ paths, train for 500,000 updates on a training set consisting of $2^{19}$ paths with early stopping to prevent overfitting, and finally test the algorithms on 10 different test sets consisting of $2^{17}$ paths each and use the average RSQP and its standard deviation as a performance measure. The different policies and value functions used by the DRL algorithms are approximated using feed-forward neural networks.

\subsubsection{States}

The states input into the DRL algorithms at each time step~$t$ consist of three variables: the current normalized time step~$\frac{t}{T}\in(0,1)$, the current normalized underlying stock price~$\frac{S_t}{S_0}\in\mathbb{R}^+$, and the current normalized hedging portfolio value~$\frac{V_t}{V_0}\in\mathbb{R}$.

\subsubsection{Actions} 

The actions at each time step $t$ consist of the next number of the underlying shares to hold $X_{t+1}\in\mathbb{R}$. However, as the DQL family of algorithms can only handle discrete action spaces (see Table~\ref{tab:DRL_algorithms}), the actions for DRL are discretized as $X_{t+1}\in[0.00, 0.02, \dots, 0.98, 1.00]$, for a total of 51 possible actions. 

\subsubsection{Episodes} 

\begin{figure*}[ht]
    \fbox{\parbox{\textwidth}{
    \centering
    \includegraphics[width=1.0\linewidth]{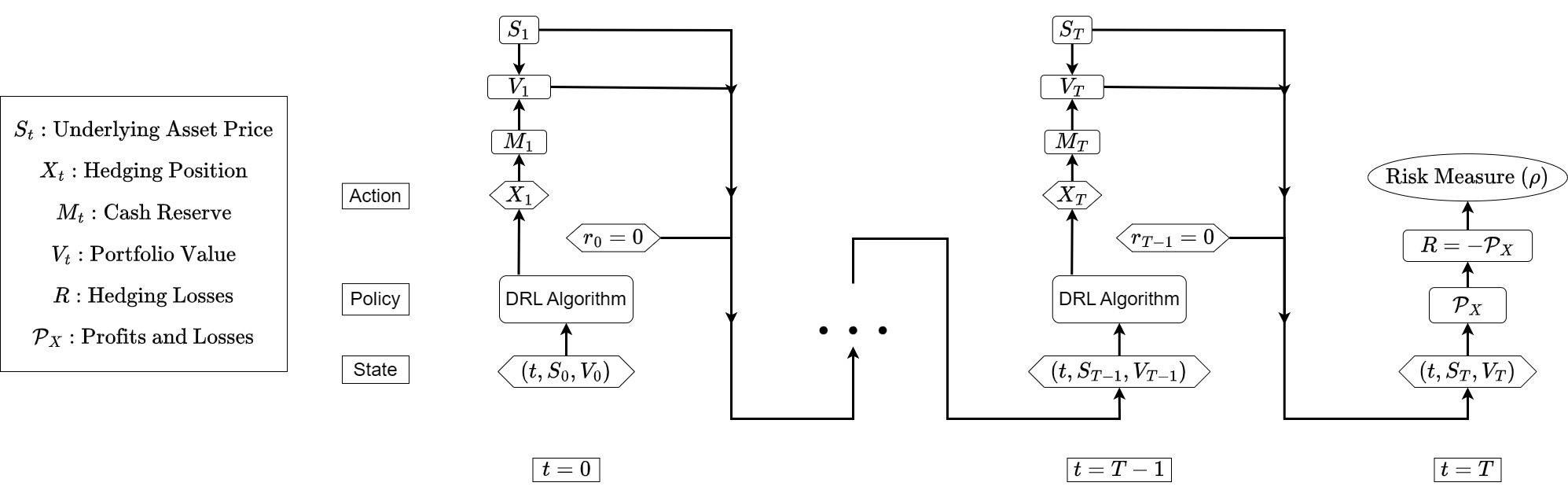}
    % \includesvg[width=1.0\linewidth]{new_architecture.svg}
    }}
    \caption{An episode of DRL dynamic hedging from $t=0,\dots,T$}
    \label{fig:architecture}
\end{figure*}
Figure~\ref{fig:architecture} depicts an episode of DRL dynamic hedging for each time step $t=0,\dots,T$. At each time step, the current state $s_t$ is fed into the DRL algorithm which then outputs the action $X_{t+1}$. This is repeated until the final time step where the hedging loss $R$ is computed. During the training of MCPG and PPO, which are Monte Carlo algorithms and therefore rely on entire episodes to update their policy, the reward is backpropagated through time. Therefore, the neural network approximators used for the policy contain an inherent recurrent connection. This is not the case for the value function neural network approximators which are updated through Temporal-Difference (TD) learning.

\subsubsection{Rewards}

The only reward in an episode is the negative of the asymmetric squared hedging loss at the last stage $T$: ${r_T=-R^2\mathds{1}_{ \{R>0\} }}$ found in Eq.~\ref{eq:RSQP}. We take the negative of the squared loss since maximizing rewards entails minimizing risk. The expected reward is estimated empirically from a mini-batch of size $N$: 
\begin{equation}
    \mathbb{E}[r_T]=-\sqrt{\frac{1}{N}\sum^N_{n=1}R_n^2\mathds{1}_{\{R_n>0\}}}.    
\end{equation}
%with $R = -\mathcal{P}_X$ from Eq.~\ref{eq:profits}. 
Rewards are null for all prior time steps, i.e. $r_t = 0$ for $t=0,\dots,T-1$. Rewards in this environment are therefore sparse, with a single reward being provided on each episode.

\subsubsection{Early Stopping}

Early stopping was implemented by training the DRL algorithms on the training set and testing them on the validation set every 1000 updates. 
For any given algorithm, we train it until two conditions are met: 1) the last 5 logged validation RSQPs are higher than the 6th last validation RSQP, and 2) these 6 validation RSQPs are lower than that of the B-S DH baseline (see Section~\ref{se:bsdh}). 
Additionally, we save the model achieving the lowest validation RSQP throughout the training iterations and use that model to compute the out-of-sample performance of the algorithm on the test set.
% For any given algorithm, if it outperformed the B-S DH baseline (see Section~\ref{se:bsdh}) by achieving a lower RSQP on the validation set, we logged the number of training iterations at which the smallest RSQP was attained on the validation set. We then trained that DRL algorithm on the training set only up to that number of iterations. Conversely, if the algorithm did not outperform the baseline, no early stopping was implemented. 
Notably, only the MCPG algorithm triggered early stopping.

\section{Experimental Results}

\begin{table}[ht]
    \centering
    \begin{tabular}{l|l|p{0.7cm}|p{1cm}}
        \toprule
        \toprule
        Algorithm & Average RSQP & $p$-value & Runtime\footnotemark{} (hh:mm)\\
        \midrule
        \midrule
        \textbf{MCPG} & $\mathbf{0.8111 \ (0.0210)}$ & 0.00 &$00:24$\\
        \midrule
        B-S DH (Baseline) & $0.9038 \ (0.0074)$ & 0.00 & $00:00$\\
        \midrule
        PPO & $0.9439 \ (0.0158)$ & 0.00 & $05:58$\\
        TD3 & $1.0113 \ (0.0223)$ & 0.04 & $10:20$\\
        DQL & $1.0278 \ (0.0119)$ & 0.00 & $06:32$\\
        DDPG & $1.0467 \ (0.0089)$ & 0.00 & $09:37$\\
        Dueling DQL* & $1.0745 \ (0.0095)$ & 0.00 & $06:07$\\
        DD DQL* & $1.1111 \ (0.0109)$ & 0.00 & $06:23$\\
        Double DQL & $1.1791 \ (0.0096)$ &  & $07:34$ \\
        \bottomrule
    \end{tabular}
    \caption{Performance of DRL algorithms. Column 2: Average RSQP attained over 10 test sets (standard deviation in parenthesis). Column 3: P-values from t-tests assessing whether the mean RSQP of each algorithm is significantly lower than that of the algorithm listed directly below it. Column 4: Training time over the training set. The algorithms indicated by stars are the two variants not previously explored in the dynamic hedging literature.}
    \label{tab:Performance}
\end{table}

Table~\ref{tab:Performance} shows the performance of the eight algorithms we tested. The best performing DRL algorithm in terms of RSQP is Monte Carlo Policy Gradient (MCPG) (RSQP: $0.8111$), followed by Proximal Policy Optimization (PPO) ($0.9439$). This can be explained by the fact that MCPG and PPO are \textit{Monte Carlo (MC)} DRL algorithms, meaning that they wait until the end of an episode to perform an update step. These algorithms exhibit higher performance in the context of option hedging than \textit{Temporal-Difference (TD) learning} DRL algorithms, such as DQL and DDPG (and their variants), which perform an update at each time step. This is because the formulation of the option hedging problem leads to an environment in which rewards are sparse, where only a single reward is obtained from the environment at the final time step. Thus, early-iteration updates of TD learning algorithms are imprecise since the value function component of the target, which is its main driver when rewards are frequently null, are heavily biased at the onset.
%noisier since most of the time the rewards are 0, and early-stage updates are heavily biased due to imprecise estimates of the value function in the target.
Ultimately, MCPG is the only algorithm that was capable of significantly outperforming the B-S DH baseline in the allotted computational budget ($0.8111$ vs $0.9038$). This improvement is statistically significant at the $1\%$ confidence level as indicated by the $p$-value of 0.00 from Table~\ref{tab:Performance}.

Another advantage of MCPG is the training time. As shown in Table~\ref{tab:Performance}, the MCPG algorithm was able to converge in just over 24 minutes\footnotemark[1] after early stopping at 69,000 update steps while all other models ran for the whole 500,000 update steps and took between 5 and 10 hours to complete. The Black-Scholes delta hedge strategy remains the most efficient to compute, finishing virtually instantly since it has a closed-form solution (Eq.~\ref{eq:BSDH}).
\footnotetext{All experiments were ran on an Nvidia A100 GPU.}
%The success of PG also seems to be attributed to the fact that it learns exclusively the policy without relying on a value function. Indeed, all other DRL algorithms rely on some version of a value function: DRL and DDPG on the state-action value function and PPO on the advantage function which is computed using the state-value function. The updates to the value functions estimates rely on the rewards at each time step which are sparse, leading to noisy updates. It would seem that trying to estimate a value function in an environment with sparse rewards such as in the context of option hedging only serves to hinder the performance of those DRL algorithms and might also explain why even a Monte Carlo DRL algorithm such as PPO seems to struggle to outperform the baseline.

\paragraph{DQL algorithms.}

Although DQL is not able to outperform the B-S DH baseline ($1.0278$ vs $0.9038$), as shown in Table~\ref{tab:Performance}, its performance is between that of TD3 (1.0113) and DDPG (1.0467). We can also see that the performance of Dueling DQL is marginally worse ($1.0745$) compared to DQL, while the performance of Double DQL is significantly worse ($1.1791$). This seems to indicate that trying to identify states where actions have little impact by implementing Dueling DQL does not provide an advantage, while trying to fix the overestimation problem frequently encountered in DQL by implementing Double DQL hinders performance. Additionally, combining the two methods by implementing Dueling Double DQL (DD DQL) attains an RSQP of $1.1111$, indicating a performance that is between that of Dueling DQL and Double DQL. Notably, variants of DQL achieve the lowest performance out of the 8 DRL algorithms explored, except for vanilla DQL. This can be attributed to the simplicity of DQL compared to its more complex variants, which is an advantage when computational resources are limited. Indeed, vanilla DQL must only learn the state-action value function, approximated using a single neural network, whereas Dueling DQL must learn the state value function and the advantage function using the same neural network. The Double DQL architecture requires optimizing over two neural networks, each representing a different state-action value function, which further increases the complexity of the algorithm.

\paragraph{DDPG algorithms.}

DDPG performs marginally worse than DQL ($1.0467$ vs $1.0278$), which is surprising because it has the advantage over DQL of naturally being able to handle continuous action spaces. Therefore, the precision of the action outputs does not suffer from discretization and the hedging strategy can be more refined. The complexity of DDPG could be a factor of its under-performance compared to DQL. Indeed, DDPG learns both a policy and a state-action value function, approximated using two distinct neural networks. This requires not only training an additional neural network but also tuning an additional 5 hyperparameters as seen from Table~\ref{tab:Hyperparameters}.

Twin-Delayed DDPG (TD3) seeks to address the overestimation problem typically encountered in DDPG by taking the minimal state-action value between two state-action value functions. This seems to only marginally lower the RSQP over vanilla DDPG ($1.0113$ vs $1.0467$), indicating that once again the overestimation problem is not very severe in our environment. Note, however, that TD3 outperforms DQL ($1.0113$ vs $1.0278$), whereas DDPG is unable to do so. Nonetheless, judging from the $p$-value of 0.04 from Table~\ref{tab:Performance}, the performance improvement of TD3 over DQL is only statistically significant at the 5\% confidence level. Furthermore, this performance gain of TD3 over DQL comes at the cost of taking significantly more time to train (10:20 vs 6:32), while still being unable to outperform the baseline's RSQP of $0.9038$.

\subsubsection{Hyperparameters}

As indicated in Section~\ref{se:DRL_Setting}, the function approximators used for the value functions and policy are feed-forward neural networks for which we selected the hyperparameters among the following values through grid search: learning rates in [0.001, 0.0001, 0.00001], batch sizes in [64, 128, 256], number of hidden layers in [2, 3, 4], and hidden layer sizes in [64, 128, 256]. Thus $3^4=81$ combinations of hyperparameters were tested. 

Interestingly, during hyperparameter tuning, 80 out of 81 combinations of hyperparameters lead to a validation RSQP that is lower than the baseline for MCPG, with values ranging from 0.8142 to 0.9084. We can conclude that MCPG is robust to the choice of hyperparameters. For PPO, only three combinations of hyperparameters were able to outperform the baseline during hyperparameter tuning, with RSQPs of 0.8947, 0.8969, and 0.9032. However, the validation losses over time showed that training was very unstable for these three best PPO hyperparameter combinations with large swings in the RSQP. Therefore, Table~\ref{tab:Hyperparameters} is based on the 4th best hyperparameter combination which had a smooth and stable training curve, but a validation RSQP of 0.9333. As such, this architecture was still unable to outperform the baseline. No hyperparameter combination for the other algorithms was able to outperform the baseline within the provided computational budget.

\begin{table}[ht]
    \centering
    \begin{tabular}{l|r|r|r|r}
        \toprule
        \toprule
        Algorithm & Learning & Batch & \# Hidden & Hidden\\
        & Rate & Size & Layers & Size\\
        \midrule
        MCPG & 0.00001 & 256 & 4 & 64\\
        PPO & 0.00001 & 128 & 2 & 256\\
        TD3 & 0.00001 & 64 & 4 & 256\\
        DQL & 0.00010 & 64 & 4 & 128\\
        DDPG & 0.00001 & 64 & 4 & 256\\
        Dueling DQL & 0.00010 & 128 & 3 & 256\\
        DD DQL & 0.00010 & 256 & 3 & 256\\
        Double DQL & 0.00010 & 64 & 4 & 64 \\
        \bottomrule
    \end{tabular}
    \caption{Optimal hyperparameters of the DRL algorithms neural networks chosen over 81 combinations.}
    \label{tab:Hyperparameters}
\end{table}

Table~\ref{tab:Hyperparameters} shows the hyperparameters resulting in the lowest RSQP on the validation set for the neural networks of the DRL algorithms studied. These were also the hyperparameters which achieved the results from Table~\ref{tab:Performance}.

%%%%%%%%%%%%%%%%%%%%%%%%%%%%%%

\begin{figure*}[ht]
    \centering
    \def\svgwidth{1.0\linewidth}
    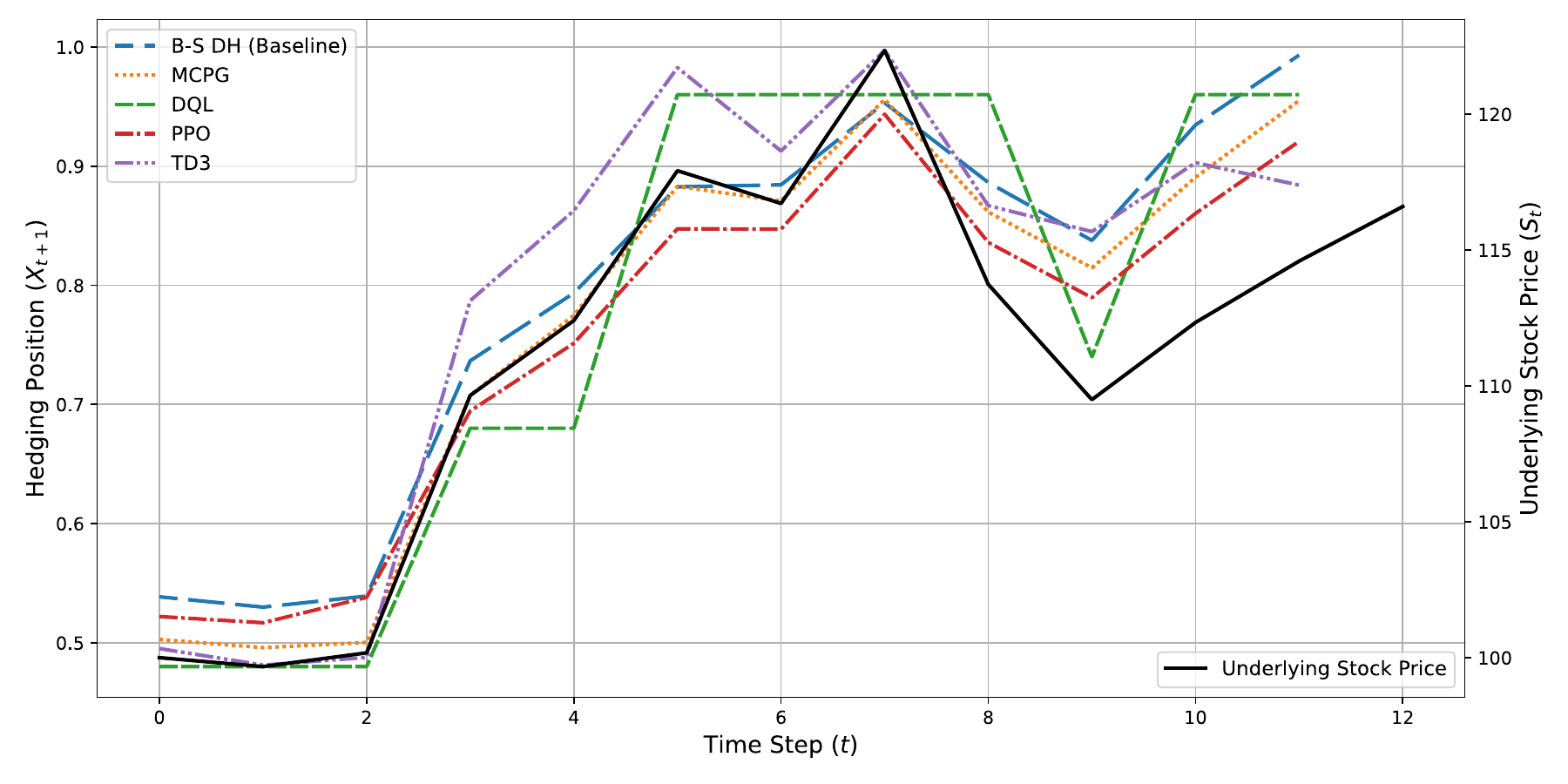
    \caption{Hedging position $\{X_{t+1}\}^{T-1}_{t=0}$ and underlying stock price $\{S_t\}^{T}_{t=0}$ for one simulated episode.}
    \label{fig:strategy}
\end{figure*}

\subsubsection{Hedging Strategy}

Figure~\ref{fig:strategy} depicts a single simulated underlying stock price path $\{S_t\}^T_{t=0}$ following Eq.~\ref{eq:gjr-garch}, along with the sequence of hedging positions $\{X_{t+1}\}^{T-1}_{t=0}$ obtained from the best performing variants of the DRL algorithms of Table~\ref{tab:Performance} (namely: MCPG, DQL, PPO and TD3) and for the B-S DH baseline. As the figure shows, the hedging positions of the algorithms and the baseline are influenced by the movements of the underlying stock price. The best performing DRL algorithms (MCPG and PPO) closely follow each other and are near the B-S DH baseline strategy. Conversely, DQL and TD3 positions do not seem to closely align, reflecting poor training outcomes, which explain the poor performance on the test set.

%Dueling DQL seems to struggle to refine its hedging strategy as indicated by the fact that its hedging strategy varies by a large amount of shares from one time step to the next instead of the minimum 0.05 difference possible from the discretized actions. This leads to poor risk minimization which is also indicated by the RSQP from Table~\ref{tab:Performance} which is higher than that of the baseline.

%Although the hedging strategy output by TD3 is more refined than Dueling DQL on account of it being able to output continuous actions, it still struggles to output actions which are near-optimal as defined by the hedging strategy output by PG. As can be observed in Figure~\ref{fig:strategy}, the hedging strategy diverges from the near-optimal actions of PG.

\subsubsection{Limitations}

The main limitation common to most DRL algorithms and also present in our work is their heavy computational load coupled with their dependence on hyperparameters, forcing the hyperparameter search to be narrow. Different hyperparameters might improve performance, perhaps enough to allow outperforming the baseline, whereas most algorithms studied in this work are not currently able to do so. 

Some of the algorithms have additional hyperparameters that we were unable to optimize due to computational limitations. Examples of these hyperparameters include, but are not limited to, the target network learning rate $\Bar{\beta}$ from DQL, which is used to update the target network parameters, $\epsilon$ from PPO, which determines the range within which the policy updates are clipped, and both learning rates $\alpha$ and $\beta$ from PPO and DDPG, which could have different values but are chosen to be the same due to limited computational resources in our case.
As Table~\ref{tab:DRL_algorithms} shows, MCPG has only 9 hyperparameters to tune whereas the next DRL algorithms with the lowest number of hyperparameters are DQL and PPO with almost twice that of MCPG. More powerful computational resources would allow us to explore these additional hyperparameters and most likely lead to a better performance.
%\newpage

\section{Conclusion and Further Work}

This paper contributes to the field of DRL for dynamic option hedging by \textbf{1)} providing a benchmark for a variety of DRL algorithms for the task of dynamic hedging. Where previous work mostly focuses on the performance of one or two algorithms at a time, we compare the performance of eight DRL algorithms; thus providing a more holistic comparison. \mbox{\textbf{2)} This} paper explores the performance of two variants, which, as far as we know, have never been studied for the task of dynamic hedging; Dueling DQL and Dueling Double DQL.

Our work shows that the MCPG algorithm performs the best in terms of risk reduction as measured by the RSQP, closely followed by PPO. However, only the MCPG algorithm is able to outperform the Black-Scholes delta hedge baseline strategy, whereas PPO and all variants of DQL and DDPG fail to do so. 
% Nevertheless, Dueling DQL, Dueling Double DQL and TD3 all manage to outperform their vanilla variants. 
Additionally, the training time of MCPG is more than fourteen times smaller than the closest competitor. Our findings show that the MCPG algorithm inspired by \cite{BuehlerDeepHedging} should be the preferred approach when tackling deep hedging problems in practice.

%We also graph the hedging strategy output by each algorithm for a single simulated path of the underlying stock to analyze the behaviour that causes an algorithm to outperform or underperform over the baseline.

One reason why strategies output by algorithms employing a value function perform poorly is that the considered hedging task is framed as a sparse reward problem. In further work, we could contemplate the approach used for instance in \cite{ChongVariable} which consists in breaking down the single reward into subparts to create dense rewards; assessing whether this approach improves the performance of value-based RL algorithms would be relevant.  

%It would also be interesting to find a way to decompose the RSQP risk measure used such that the dynamic hedging environment outputs a reward at each time step (dense rewards) as opposed to a single risk measure at the end of an episode (sparse rewards). This should increase the performance of the algorithms approximating value functions (DQL, PPO, and DDPG).
%We could also experiment with higher-dimensional problems to see if the more sophisticated DRL algorithms outperform PG, such as portfolio optimization or dynamic hedging using multiple hedging instruments instead of only using the underlying stock price. 
Moreover, the option hedging environment used in this study is low-dimensional since we only hedge the sale of a call option using a single hedging instrument (i.e. the underlying stock). Therefore, we have a limited number of state variables and the actions are one-dimensional. Other research works focus on hedging the sale of an option using multiple other hedging instruments (e.g. multiple different option contracts such as in \cite{CarbonneauLongTerm,CarbonneauGodin2021DeepEqualRiskPricingMultiple}) in richer state spaces, where the number of inputs and outputs can increase dramatically. Analyzing whether the outperformance of MCPG over the other algorithms persists in higher-dimensional tasks would be interesting future work.

Finally, implementing the different algorithms studied in this paper for other financial sequential decision-making tasks such as portfolio optimization and optimal execution would allow assessing whether alternative shapes of the value function associated with these tasks could impact the performance of value-based algorithms. 

%\textbf{ The under-performance of DQL incomparison to PG is expected to be even worse is such high-dimensional environments. However it is unclear} %In these high-dimensional environments, we might see a relatively simple DRL algorithm such as PG under-perform compared to more complex actor-critic algorithms such as PPO and DDPG. However, note that an increase in performance for DQL is not expected since it is known to under-perform in high-dimensional environments.

%% The file named.bst is a bibliography style file for BibTeX 0.99c
\bibliographystyle{named}
\bibliography{mainArxiv}

\end{document}